\documentclass[conference, 10pt]{IEEEtran}

\usepackage{cite}
\usepackage{url}
\usepackage{graphicx}
\usepackage{color}
\usepackage{placeins}
\usepackage{float}
\usepackage{tabularx,colortbl}
\usepackage{ifthen}
\usepackage{amssymb}
\usepackage{amsmath}

\usepackage{tikz}
\usepackage{pgfplots}
\pgfplotsset{scaled y ticks=false}
\usepackage{mathrsfs}
\usepackage{mathtools}
\usetikzlibrary{arrows}
\usepgfplotslibrary{external}
\usepackage{xcolor}

\usepackage{pifont}
\pgfplotsset{grid style={dashed,gray}}
\pgfplotsset{minor grid style={dotted,gray}}
\pgfplotsset{major grid style={dashed,gray}}

\makeatletter

\newcounter{author}
\renewcommand{\author}[2][]{
   \stepcounter{author}
   \@namedef{author@\theauthor}{#2}
   \@namedef{authorlabel@\theauthor}{#1}
}

\newcounter{address}
\newcommand{\address}[2][]{
   \stepcounter{address}
   \@namedef{address@\theaddress}{#2}
   \@namedef{addresslabel@\theaddress}{#1}
}

\newcommand{\alsep}{and}

\def\newmaketitle{\par%
  \begingroup%
  \normalfont%
  \def\thefootnote{}
  \def\footnotemark{}
  \let\@makefnmark\relax
  \footnotesize
  \footnotesep 0.7\baselineskip
  \normalsize%
  \twocolumn[\thenewmaketitle\@IEEEaftertitletext]%
  \if@IEEEusingpubid
     \enlargethispage{-\@IEEEpubidpullup}%
  \fi
  \endgroup
  \setcounter{footnote}{0}\let\maketitle\relax\let\@maketitle\relax
  \gdef\@thanks{}%
  \let\thanks\relax}

\def\thenewmaketitle{
  \newpage
  \begin{center}%
    \vskip0.2em{\Huge\@IEEEcompsoconly{\sffamily}\@IEEEcompsocconfonly{\normalfont\normalsize\vskip 2\@IEEEnormalsizeunitybaselineskip
   \bfseries\large}\@title\par}\vskip1.0em\par%
    \vspace{1ex}
    \newcounter{c@author}
    \newcounter{c@tmp}
    \ifthenelse{\value{author}=2}{%
      \newcommand{\liand}{ and }}{%
      \newcommand{\liand}{, and }}
    \ifthenelse{\value{address}<2}{%
      \@nameuse{author@1}%
      \stepcounter{c@author}%
      \whiledo{\value{c@author}<\value{author}}{%
        \setcounter{c@tmp}{\value{author}}%
        \addtocounter{c@tmp}{-\value{c@author}}%
        \ifthenelse{\value{c@tmp}=1}{%
          \renewcommand{\alsep}{\liand}}{\renewcommand{\alsep}{, }}%
        \stepcounter{c@author}\alsep \@nameuse{author@\thec@author}}\\%
    }
    {
      \@nameuse{author@1}${}^{(\ref{\@nameuse{authorlabel@1}})}$%
      \stepcounter{c@author}%
      \whiledo{\value{c@author}<\value{author}}{%
      \setcounter{c@tmp}{\value{author}}%
      \addtocounter{c@tmp}{-\value{c@author}}%
      \ifthenelse{\value{c@tmp}=1}{%
        \renewcommand{\alsep}{\liand}}{\renewcommand{\alsep}{, }}%
      \stepcounter{c@author}\alsep \@nameuse{author@\thec@author}%
        ${}^{(\ref{\@nameuse{authorlabel@\thec@author}})}$%
      }
    }
    \vspace{0.2ex}

    \ifthenelse{\value{address}>0}{%
      \ifthenelse{\value{address}=1}{
        {\@nameuse{address@1}}
      }
      {
        \newcounter{c@address}

        \begin{center}
        \whiledo{\value{c@address}<\value{address}}
        {
          \refstepcounter{c@address}
            ${}^{(\thec@address)}$\,%
              \label{\@nameuse{addresslabel@\thec@address}}%
              \@nameuse{address@\thec@address}\\ %
        }
        \end{center}
      } 
    }
    {
      \relax
    }
  \end{center}
}

\makeatother

\usepackage{acronym}
\newacro{5g} [5G] {fifth-generation}
\newacro{6g} [6G] {sixth-generation}
\newacro{isac} [ISAC] {integrated sensing and communication}
\newacro{ul} [UL] {uplink}
\newacro{mu} [MU] {multi-user}
\newacro{csi} [CSI] {channel state information}

\newacro{em} [EM] {electromagnetic}
\newacro{siso} [SISO] {single input single output}
\newacro{miso} [MISO] {multiple input single output}
\newacro{mimo} [MIMO] {multiple input multiple output}
\newacro{xl-mimo} [XL-MISO] {extremely large-scale MIMO}
\newacro{ras} [RAs] {reconfigurable antennas}
\newacro{los} [LoS] {line-of-sight}
\newacro{nlos} [NLoS] {non-line-of-sight}
\newacro{shod} [SHOD] {spherical harmonious orthogonal decomposition}
\newacro{ofdm} [OFDM] {orthogonal frequency division multiplexing}
\newacro{dof} [DoF] {degree of freedom}
\newacro{fim} [FIM] {Fisher information matrix}
\newacro{em} [EM] {electromagnetics}
\newacro{er-fas} [ER-FAS] {electromagnetically reconfigurable FAS}
\newacro{eras} [ERAs] {electromagnetically reconfigurable antennas}
\newacro{aod} [AOD] {angle-of-departure}
\newacro{aoa} [AOA] {angle-of-arrival}
\newacro{aoas} [AOAs] {angles-of-arrival}
\newacro{sp} [SP] {scatter point}
\newacro{ml} [ML] {maximum likelihood}
\newacro{mse} [MSE] {mean square error}
\newacro{snr} [SNR] {signal-to-noise ratio}
\newacro{lmr} [LMR] {line-of-sight to multipath ratio}

\newacro{rmse} [RMSE] {root mean square error}
\newacro{crb} [CRB] {Cram\'er-Rao bound}
\newacro{peb} [PEB] {position error bound}
\newacro{kld} [KLD] {Kullback–Leibler divergence}
\newacro{siso} [SISO] {single-input-single-output}
\newacro{mimo} [MIMO] {multiple-input multiple-output}
\newacro{mcrb} [MCRB] {misspecified Cram\'er-Rao bound}
\newacro{bs} [BS] {base station}
\newacro{ue} [UE] {user equipment}
\newacro{arv} [ARV] {array response vector}
\newacro{upa} [UPA] {uniform planar array}
\newacro{rf} [RF] {radio frequency}
\newacro{bb} [BB] {baseband}
\newacro{ris} [RIS] {reconfigurable intelligent surface}
\newacro{rms} [RMS] {root mean square}
\newacro{psd} [PSD] {positive semidefinite}
\newacro{lmi} [LMI] {linear matrix inequality}
\newacro{bca} [BCA] {block-coordinate ascent}
\newacro{bcd} [BCD] {block-coordinate descent}

\newacro{ma} [MA] {movable antenna}
\newacro{6dma} [6DMA] {six-dimensional movable antenna}
\newacro{pra} [PRA] {pattern reconfigurable antenna}
\newacro{ra} [RA] {reconfigurable antenna}
\newacro{pra_p} [PRAs] {pattern reconfigurable antennas}
\newacro{espar} [ESPAR] {electronically steerable parasitic array radiator}
\newacro{music} [MUSIC] {multiple signal classification}
\newacro{rss} [RSS] {received signal strength}
\newacro{toa} [TOA] {time of arrival}
\newacro{wsn} [WSN] {wireless sensor network}
\newacro{nm} [NM] {Nelder-Mead}
\newacro{ls} [LS] {least-squares}
\newacro{sdp} [SDP] {semidefinite program}
\newacro{iot} [IoT] {internet of things}

\newacro{mpc} [MPC] {multipath component}
\newacro{rcs} [RCS] {radar cross section}

\newacro{fas} [FAS] {fluid antenna system}
\newacro{fas_p} [FAS] {fluid antenna systems}
\newacro{sr-fas} [SR-FAS] {spatially reconfigurable FAS}
\newacro{ris} [RIS] {reconfigurable intelligent surface}
\newacro{omp} [OMP] {orthogonal matching pursuit}
\newacro{ci} [CI] {confidence interval}
\newacro{sage} [SAGE] {space-alternating generalized expectation-maximization}

\newacro{scsi} [sCSI] {spatial domain channel state information}
\newacro{ecsi} [eCSI] {EM domain channel state information}

\newacro{scm} [SCM] {sample covariance matrix}
\newacro{evd} [EVD] {eigen-value decomposition}
\newacro{dof} [DoF] {degree of freedom}

\allowdisplaybreaks

\title{Low-Complexity Channel Estimation for Reconfigurable Fluid Antenna System}

\author[org1]{Alireza Fadakar}
\author[org1]{Andreas F. Molisch}

\address[org1]{Ming Hsieh Department of Electrical and Computer Engineering, University of Southern California, Los Angeles, California, USA, \{fadakarg, molisch\}@usc.edu}

\usepackage{fancyhdr}
\fancypagestyle{arxivnotice}{
  \fancyhf{} 
  
  \fancyhead[C]{\small \textit{This paper has been accepted for publication in the 2026 IEEE International Symposium on Antennas and Propagation and USNC-URSI Radio Science Meeting (AP-S/URSI).}}
  
  \fancyfoot[C]{\scriptsize \copyright~2026 IEEE. Personal use of this material is permitted. Permission from IEEE must be obtained for all other uses, in any current or future media, including reprinting/republishing this material for advertising or promotional purposes, creating new collective works, for resale or redistribution to servers or lists, or reuse of any copyrighted component of this work in other works.}
}

\begin{document}

\newmaketitle
\thispagestyle{arxivnotice}

\begin{abstract}
Electromagnetically reconfigurable fluid antenna system (ER-FAS) provides additional electromagnetic (EM)-domain degrees of freedom by enabling dynamic control of per-element radiation patterns, thereby improving power efficiency in wireless communications. The effective exploitation of these capabilities, however, critically depends on accurate channel estimation, which has received limited attention in prior studies. This paper presents a low-complexity channel parameter estimation framework for downlink wideband systems that leverages available channel sparsity. A synthesis-based reconfigurability model is considered, where each antenna generates desired beampatterns using a finite set of EM basis functions. Based on this model, a joint optimization of digital and EM precoders is formulated with the objective of minimizing the Cram\'er-Rao lower bound of the channel parameters. 
Simulation results demonstrate that the proposed hybrid beamforming and estimation approach enables accurate recovery of channel parameters while maintaining low computational complexity.
\end{abstract}


\bstctlcite{IEEEexample:BSTcontrol}
\acresetall

\section{Introduction}
The transition to \ac{5g} and \ac{6g} networks requires substantial gains in data rate and capacity, motivating new physical-layer solutions to overcome inherent performance limits \cite{wang2025RA,Wong2023fluid, fadakar2025stacked, Ashkan2026race_cma}. 
Enhancing \ac{bs} capabilities is therefore essential, as overall system throughput is constrained by the efficient utilization of limited physical resources \cite{Ying2024Reconfigurable, Fadakar2026Hybrid,Ashkan2025CL_ISAC,Chen2023_5G_6G}. 
While time-frequency optimization is well established, spatial-domain techniques, particularly \ac{mimo}, introduce additional \ac{dof}. 
However, further aperture scaling in mast-mounted deployments is restricted by structural (form-factor) constraints.

\Acp{fas} have emerged as a promising physical-layer technology for \ac{6g} and beyond by enabling dynamic reconfiguration of the radiating structure to exploit favorable spatial channel conditions \cite{Wong2023fluid}. 
Unlike conventional fixed arrays, \acp{fas} provide additional spatial-diversity gains through adaptive antenna relocation or pattern reshaping.
\Acp{fas} can be broadly classified into two categories: 
\ac{sr-fas} and \ac{er-fas}. 
The former relies on port selection to reposition the radiating element while preserving its intrinsic \ac{em} characteristics \cite{New2024fluid_Tutorial}. 
In contrast, \ac{er-fas}, also known as \acp{ra}, maintain a fixed physical location and dynamically modify their metallic or dielectric structures to control operating frequency, polarization, and radiation pattern \cite{wang2025RA,Rodrigo2014frequency}. 
This work focuses on radiation-pattern-based \ac{er-fas}, whose radiation characteristics can be reconfigured in real time through controlled manipulation of the fluidic \ac{rf} radiator.

The effective utilization of \acp{er-fas} critically depends on accurate acquisition of \ac{ecsi}, which has received comparatively limited attention in existing studies.
Several works have addressed channel estimation challenges in \ac{er-fas}-assisted uplink \ac{mimo} systems \cite{Ying2024Reconfigurable,Bahceci2017csi,Liang2024deep_csi}. 
Unlike conventional \ac{mimo} systems that require estimation of a single channel matrix, \ac{er-fas}-enabled systems may exhibit multiple channel realizations corresponding to different radiation patterns, resulting in substantial pilot overhead \cite{Ying2024Reconfigurable}. 
To mitigate this challenge, \cite{Bahceci2017csi} proposed a Gram-Schmidt-based decomposition of radiation patterns into orthogonal basis functions, effectively decoupling antenna characteristics from the propagation channel and reducing channel estimation overhead. 
In contrast, \cite{Liang2024deep_csi} introduced a deep learning-based channel extrapolation approach, where a neural network infers channels corresponding to unobserved radiation patterns from limited pilot measurements. 
More recently, \cite{Ying2024Reconfigurable} developed subspace-based channel estimation methods, including wideband \ac{scsi} estimation followed by parameterized \ac{aoa} extraction and \ac{ls}-based reconstruction of the full \ac{ecsi}.

Despite the aforementioned progress in \ac{er-fas}-based channel estimation, several challenges persist. 
Most existing methods focus on uplink \ac{mimo} scenarios and do not readily generalize to downlink \ac{miso} systems. 
Moreover, the approaches in \cite{Bahceci2017csi,Liang2024deep_csi} are limited to a finite set of predefined radiation patterns, making them unsuitable for channel inference over a continuous pattern space. Subspace-based methods \cite{Ying2024Reconfigurable} further require a large number of pilots for accurate \ac{scm} estimation and rely on \ac{evd}, leading to increased overhead and computational complexity.

Motivated by these gaps, this paper investigates low-complexity \ac{ecsi} estimation for \ac{er-fas}. The main contributions are summarized as follows:
\begin{itemize}
\item
A synthesis-based signal model for \ac{er-fas}-assisted channel estimation is developed, where each antenna element generates beampatterns from a finite set of basis functions. 
Based on this model, closed-form expressions for the \ac{fim} and the corresponding \acp{crb} are derived, establishing fundamental performance limits.
\item
Leveraging the derived bounds, a joint optimization framework for the digital and \ac{em} precoders is formulated to minimize the \ac{crb} of the channel parameters, resulting in a computationally efficient codebook-based design.
\item
A low-complexity channel estimation method exploiting channel sparsity is proposed and evaluated via simulations, demonstrating close agreement with the \ac{crb} and significant channel parameter performance gains enabled by \ac{er-fas}, compared to non-reconfigurable arrays.
\end{itemize}

\vspace{-0.2cm}
\section{System and Signal Model}
\subsection{System Model}
\begin{figure}
\centering
\includegraphics[width=0.7\columnwidth]{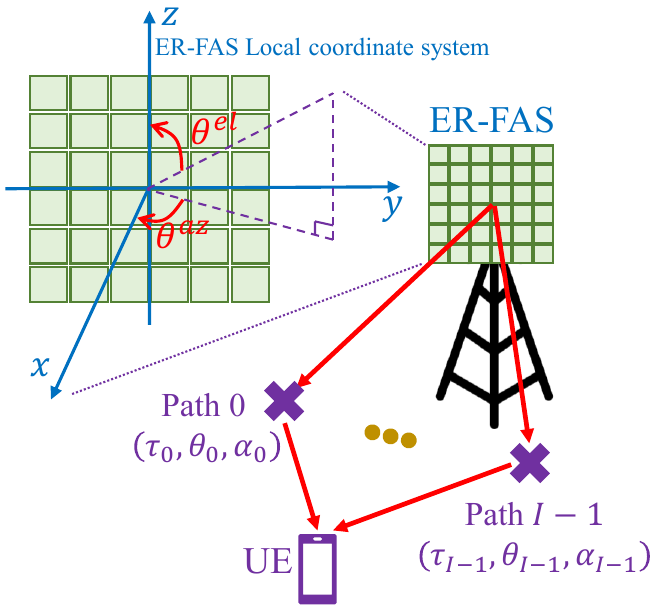}
\vspace{0.1cm}
\caption{
Downlink \ac{miso} system model employing an \ac{er-fas} for channel parameter estimation.
}
\label{fig:system-model}
\end{figure}

As illustrated in Fig.~\ref{fig:system-model}, we consider a downlink \ac{miso} communication scenario in which the \ac{bs} is equipped with a multi-element \ac{er-fas} arranged in a \ac{upa}. The array comprises $M = M_1 M_2$ radiating elements, where $M_1$ and $M_2$ denote the numbers of rows and columns, respectively. Each \ac{er-fas} element can independently synthesize a prescribed radiation pattern. The \ac{ue} is assumed to be equipped with a single isotropic antenna.

\vspace{-0.2cm}
\subsection{Signal Model}
To estimate the channel parameters, the \ac{bs} transmits $N_t$ sequential downlink \ac{ofdm} pilots over $N_s$ subcarriers. Assuming all pilot symbols are equal to $1$, the received signal at the \ac{ue} for the $t$-th transmission on the $n$-th subcarrier is given by
\begin{equation}\label{eq:y-def}
y_{t,n}
=
\sqrt{P}\,
\mathbf{h}_{t,n}^{\mathsf{T}}
\mathbf{f}_{t}
+
v_{t,n},
\end{equation}
where $P$ denotes the transmit power, and $\mathbf{f}_t \in \mathbb{C}^{M}$ is the digital precoder used by the \ac{bs} during the $t$-th transmission. To maintain a constant total transmit power across all transmissions, the precoders are normalized such that $\sum_{t=1}^{N_t} \mathbf{f}_t^{\mathsf{H}}\mathbf{f}_t = 1$. The vector $\mathbf{h}_{t,n} \in \mathbb{C}^{M}$ denotes the \ac{scsi} between the \ac{bs} and the \ac{ue} on the $n$-th subcarrier, while $v_{t,n} \sim \mathcal{CN}(0,\sigma_v^2)$ represents additive white Gaussian noise. The dependence of $\mathbf{h}_{t,n}$ on the transmission index $t$ arises from the time-varying nature of the \ac{er-fas}. 
Assuming a geometric channel model with $I$ propagation paths, $\mathbf{h}_{t,n}$ can be expressed as \cite{wang2025RA,Ying2024Reconfigurable,Fadakar2026Hybrid, Fadakar2026RA_NF}:
\begin{equation}\label{eq:ch-def}
\mathbf{h}_{t,n}
=
\sum_{i=0}^{I-1}
\alpha_{i}
e^{-j2\pi\tau_in\Delta f}
\mathbf{q}_{t}(\boldsymbol{\theta}_{i}),
\end{equation}
where $\tau_i$ and $\boldsymbol{\theta}_i=[\theta_i^\text{el},\theta_i^\text{az}]^\mathsf{T}$ denote the propagation delay and 2D-\ac{aod} of the $i$-th path, respectively with $\theta_i^\text{el},\theta_i^\text{az}$ being the elevation and azimuth \acp{aod}~\cite{fadakar2024deepdoa}, $\alpha_i = \rho_i e^{j\varphi_i}$ is the corresponding complex gain, with $\rho_i$ and $\varphi_i$ denoting its amplitude and phase. The vector $\mathbf{q}_t(\boldsymbol{\theta}) \in \mathbb{C}^{M}$ captures the combined effect of the \ac{arv} and the antenna radiation patterns at the \ac{bs} during the $t$-th transmission for a given \ac{aod} $\boldsymbol{\theta}$, and is defined as
\begin{equation}\label{eq:q-def}
\mathbf{q}_{t}(\boldsymbol{\theta})
=
\mathbf{c}_{t}(\boldsymbol{\theta})
\odot 
\mathbf{a}(\boldsymbol{\theta}),
\end{equation}
where $\mathbf{a}(\boldsymbol{\theta})\in \mathbb{C}^{M}$ is the array \ac{arv} defined as:
\begin{equation}
\mathbf{a}(\boldsymbol{\theta})
=
e^{-j2\pi\omega_1\boldsymbol{\zeta}(M_1)}
\otimes
e^{-j2\pi\omega_2\boldsymbol{\zeta}(M_2)}.
\end{equation}
Here, $\boldsymbol{\zeta}(M) = [0, \dots, M-1]^\mathsf{T}$, and $\omega_1$, $\omega_2$ are the spatial horizontal and vertical frequencies:  
\begin{equation}
\omega_1
=
d
\sin(\theta^{\text{az}})
\sin(\theta^{\text{el}})/\lambda
,\ 
\omega_2
=
d
\cos(\theta^{\text{el}})/\lambda,
\end{equation}
where $d,\lambda$, denote the antenna spacing and wavelength. 
Moreover, in \eqref{eq:q-def}, $\mathbf{c}_t(\boldsymbol{\theta}) \in \mathbb{C}^{M}$ is the complex radiation patterns of the \ac{er-fas} elements which is detailed next.
\vspace{-0.5cm}
\subsection{Reconfigurability Model}
To capture reconfigurability, the \ac{er-fas} is modeled as synthesizing on-demand beampatterns via projection onto a predefined set of orthonormal basis functions \cite{Ying2024Reconfigurable}. 
Let $\mathbf{b}(\boldsymbol{\theta}) = [b_1(\boldsymbol{\theta}), \ldots, b_Q(\boldsymbol{\theta})]^{\mathsf{T}}$ denote the corresponding basis vector of dimension $Q$.  
Accordingly, the radiation response of the $m$-th \ac{er-fas} element during the $t$-th transmission is expressed as
\begin{equation}\label{eq:gain-def}
\bigl[\mathbf{c}_{t}(\boldsymbol{\theta})\bigr]_{m}
=
\mathbf{e}_{m,t}^{\mathsf{H}}
\mathbf{b}(\boldsymbol{\theta}),
\end{equation}
where $\mathbf{e}_{m,t} \in \mathbb{C}^{Q}$ denotes the \ac{em} precoder that weights the basis functions to synthesize the desired beampattern of the $m$-th \ac{er-fas} element during the $t$-th transmission. To satisfy energy conservation, $\lVert \mathbf{e}_{m,t} \rVert^{2} = 1$ is assumed \cite{Ying2024Reconfigurable,Fadakar2026Hybrid, Fadakar2026RA_NF}.
Let the coefficient matrix $\mathbf{E}_t\!\in\!\mathbb{C}^{M\times MQ}$ be defined as
\begin{equation}\label{def:E_synthesis}
\mathbf{E}_t = \mathrm{blkdiag}\bigl\{\mathbf{e}_{1,t}^{\mathsf{H}},\,\dots,\,\mathbf{e}_{M,t}^{\mathsf{H}}\bigr\}.
\end{equation}
Next, \eqref{eq:q-def} is represented as:
\begin{equation}\label{eq:q_represent}
\mathbf{q}_{t}(\boldsymbol{\theta})
= \mathbf{E}_t\bigl(\mathbf{a}(\boldsymbol{\theta})\otimes\mathbf{b}(\boldsymbol{\theta})\bigr)
= \mathbf{E}_t\,\mathbf{g}(\boldsymbol{\theta}),
\end{equation}
where 
$
\mathbf{g}(\boldsymbol{\theta})
= \mathbf{a}(\boldsymbol{\theta})\otimes\mathbf{b}(\boldsymbol{\theta})
\;\in\;\mathbb{C}^{MQ}
$ denotes the \ac{em}-domain \ac{arv}. 
Substituting \eqref{eq:q_represent} into \eqref{eq:ch-def} and subsequently into \eqref{eq:y-def}, and stacking the $N_s$ symbols for the $t$-th transmission yields:
\begin{equation}\label{eq:y-def-simp}
\mathbf{y}_{t}
=
\sum_{i=0}^{I-1}
\sqrt{P}\,
\underbrace{\alpha_{i}\,
\mathbf{d}(\tau_i)\,
\mathbf{g}(\boldsymbol{\theta}_{i})^{\mathsf{T}}
}_{\widetilde{\mathbf{H}}_i}\,
\mathbf{w}_t
+
\mathbf{v}_{t}\,,
\end{equation}
where $\mathbf{w}_t = \mathbf{E}_t^{\mathsf{T}}\mathbf{f}_t$ denotes the hybrid precoder that jointly captures the effects of the digital and \ac{em} precoders, and $\widetilde{\mathbf{H}}_i \in \mathbb{C}^{N_s \times MQ}$ represents the \ac{ecsi} matrix associated with the $i$-th propagation path.
Furthermore, $\mathbf{d}(\tau) \in \mathbb{C}^{N_s}$ denotes the delay steering vector, whose $n$-th entry is $[\mathbf{d}(\tau)]_n = e^{-j2\pi (n-1)\Delta f \tau}$ for $n = 1, \ldots, N_s$. In this work, \ac{shod} functions are adopted as basis functions due to their simplicity \cite{Costa2010Unified}. Under the \ac{shod} framework, arbitrary radiation patterns admit an infinite spherical-harmonic expansion \cite{Ying2024Reconfigurable}, which is truncated to the first $Q$ terms in the simulations. Further details on the definition of $b_{n}(\boldsymbol{\theta})$ via \ac{shod} are provided in \cite[Appendix~1]{Fadakar2026Hybrid}.

\section{Proposed Method}\label{sec:hybrid_design}
\subsection{CRB of Channel Parameters}\label{sec:FIM_analysis}
This subsection derives the \ac{crb} of the channel parameters and adopts it as the objective for joint digital and \ac{em} precoder design.
The channel parameter vector $\boldsymbol{\eta} \in \mathbb{R}^{5I}$ is defined as:
\begin{equation}\label{def:eta}
[\boldsymbol{\eta}]_{5i+1:5i+5}
=
[\theta_i^{\text{el}},\theta_i^{\text{az}},\tau_i,\rho_i,\varphi_i]^\mathsf{T},\ 
i=0,\dots ,I-1.
\end{equation}
The entries of the $(i,j)$-th $5\times 5$ block submatrix of of the \ac{fim} $\mathbf{J}_{\boldsymbol{\eta}} \in \mathbb{R}^{5I \times 5I}$ is given by \cite{fadakar2025near, fadakar2025mutual}:
\begin{align}\label{eq:fim_def}
[\mathbf{J}_{\boldsymbol{\eta}}]_{5i+r,5j+s}
&=
\frac{2}{\sigma^2}
\sum_{t=1}^{N_t}
\Re
\bigg\{
\left(
\frac{\partial \mathbf{x}_t}{\partial \eta_{5i+r}}
\right)^\mathsf{H}
\left(
\frac{\partial \mathbf{x}_t}{\partial \eta_{5j+s}}
\right)
\bigg\}
,
\\ \notag 
&
=
\frac{2P}{\sigma^2}\,
\Re
\{
\alpha_{i,j}^{(r,s)}\,
d_{i,j}^{(r,s)}\,
\mathbf{g}^{(s)}(\boldsymbol{\theta}_j)^\mathsf{T}
\mathbf{R}\,
\mathbf{g}^{(r)}(\boldsymbol{\theta}_i)^{*}
\}
\end{align}
for $i,j\in\{0,\dots ,I-1\}$ and $r,s\in\{1,\dots ,5\}$, 
where $\mathbf{x}_t$ denotes the noise-free counterpart of \eqref{eq:y-def-simp} and $\mathbf{R}=\sum_{t=1}^{N_t}\mathbf{R}_t$ where $\mathbf{R}_t=\mathbf{w}_t\mathbf{w}_t^\mathsf{H}$. 
Moreover, define
$\alpha_{i,j}^{(r,s)} = \alpha_i^{(r)*}\alpha_j^{(s)}$ and
$d_{i,j}^{(r,s)} = \mathbf{d}^{(r)}(\tau_i)^{\mathsf{H}}\mathbf{d}^{(s)}(\tau_j)$,
where $\mathbf{d}^{(3)}(\tau_i) = \frac{\partial \mathbf{d}(\tau)}{\partial \tau}|_{\tau=\tau_i}$ and $\mathbf{d}^{(r)}(\tau_i) = \mathbf{d}(\tau_i)$ for $r \neq 3$. 
Similarly, $\mathbf{g}^{(1)}(\boldsymbol{\theta}_i) = \frac{\partial \mathbf{g}(\boldsymbol{\theta})}{\partial \theta^{\mathrm{el}}}|_{\boldsymbol{\theta}=\boldsymbol{\theta}_i}$,
$\mathbf{g}^{(2)}(\boldsymbol{\theta}_i) = \frac{\partial \mathbf{g}(\boldsymbol{\theta})}{\partial \theta^{\mathrm{az}}}|_{\boldsymbol{\theta}=\boldsymbol{\theta}_i}$, and
$\mathbf{g}^{(r)}(\boldsymbol{\theta}_i) = \mathbf{g}(\boldsymbol{\theta}_i)$ for $r > 2$. In addition, $\alpha_i^{(4)} = e^{j\varphi_i}$, $\alpha_i^{(5)} = j\alpha_i$, and $\alpha_i^{(r)} = \alpha_i$ for $r \notin \{4,5\}$.
Hence, it follows that the elements of the \ac{fim} $\mathbf{J}_{\boldsymbol{\eta}}$ depend linearly on $\mathbf{R}$. 
Consequently, the \acp{crb} of the channel parameters, given by the diagonal entries of $\mathbf{J}_{\boldsymbol{\eta}}^{-1}$, i.e., $\mathrm{CRB}_u = [\mathbf{J}_{\boldsymbol{\eta}}^{-1}]_{u,u}$ for $u = 1, \ldots, 5I$, are convex functions of $\mathbf{R}$. 
\subsection{Proposed Hybrid Beamforming}
Under the assumption of perfect knowledge of the channel parameters $\boldsymbol{\eta}$, the optimization problem is formulated as:
\begin{subequations} \label{opt_prob_represent}
\begin{align}
\underset{
\mathbf{W}
}
{\textnormal{min}}
\quad
\underset{
\mathbf{u}
}
{\textnormal{max}}
\quad
&
\delta_u.\mathrm{CRB}_u
\left(
\mathbf{R};
\boldsymbol{\eta}
\right)
\label{def:opt-prob-represent}
\\
\textnormal{s.t.} \quad
&
\mathrm{tr}\left(
\mathbf{R}
\right)=1
, \label{opt:pow_const_w}
\\
& 
\mathrm{rank}(\mathbf{R})\le N_t \, ,
\label{opt:rank_1_constraints}
\end{align}
\end{subequations}
where $\{\delta_u\}_{u=1}^{5I}$ are nonnegative weighting coefficients that account for the different physical scalings of the channel parameters (e.g., delays and \acp{aod}) and allow the designer to prioritize selected \acp{crb} in the objective function.
Since $\mathrm{CRB}_u$ is convex with respect to $\mathbf{R}$, dropping the rank constraint \eqref{opt:rank_1_constraints} yields a convex optimization problem. 
To obtain a low-complexity solution, we follow an approach similar to \cite[Appendix~B]{Fadakar2026Hybrid} and exploit the inherent structure of the optimal matrix $\mathbf{R}$, which admits the following representation:
\begin{equation}\label{eq:w_low_dim_structure}
\mathbf{R}
=
\mathbf{G}_w
\mathbf{\Xi}
\mathbf{G}_w^\mathsf{H} \,,
\end{equation}
where $\boldsymbol{\Xi} \in \mathbb{C}^{3I \times 3I}$ is a \ac{psd} matrix and $\mathbf{G}_w \in \mathbb{C}^{MQ \times 3I}$ is defined blockwise as
\begin{equation}\label{eq:opt_beams}
[\mathbf{G}_w]_{3i+1:3i+3}
=
\big[
\mathbf{g}^{(1)}(\boldsymbol{\theta}_i)^{*},
\mathbf{g}^{(2)}(\boldsymbol{\theta}_i)^{*},
\mathbf{g}^{(3)}(\boldsymbol{\theta}_i)^{*}
\big],
\end{equation}
where $i = 0, \ldots, I-1$, and $\mathbf{g}^{(r)}(\cdot)$ is defined in Sec.~\ref{sec:FIM_analysis}. By relaxing $\boldsymbol{\Xi}$ to be diagonal, we obtain $3I$ codewords $\{\mathbf{w}_s\}_{s=1}^{3I}$ that achieve an approximate optimum of \eqref{opt_prob_represent}, given by
\begin{equation}\label{eq:w_opt_synthesis}
\hat{\mathbf{w}}_s
=
\hat{\mathbf{w}}_r(\boldsymbol{\theta_i})
=
\sqrt{\varrho_s}\,
\tilde{\mathbf{g}}^{(r)}(\boldsymbol{\theta}_i)^{*},
\end{equation}
where $s = 3i + r$, $i = 0, \ldots, I-1$, and $r = 1,2,3$. Here,
$\tilde{\mathbf{g}}^{(r)}(\boldsymbol{\theta}_i)
=
\mathbf{g}^{(r)}(\boldsymbol{\theta}_i) /
\lVert \mathbf{g}^{(r)}(\boldsymbol{\theta}_i) \rVert$,
and $\varrho_s \in [0,1]$ denotes the power fraction allocated to the $s$-th codeword.

Finally, the digital and \ac{em} precoders are recovered from $\mathbf{w}_s=\mathbf{E}_s^{\mathsf{T}}\mathbf{f}_s$. 
Using the definition of $\mathbf{E}_s$ in \eqref{def:E_synthesis}, we obtain $[\mathbf{w}_s]_{(m-1)Q+1:mQ} = [\mathbf{f}_s]_m \mathbf{e}_{m,s}$ for $m=1,\ldots,M$. Substituting \eqref{eq:w_opt_synthesis} yields $[\mathbf{f}_s]_m \mathbf{e}_{m,s} = \sqrt{\varrho_s}\,\mathbf{g}^{(r)}_m(\boldsymbol{\theta}_i)^{*}$. Since $\lVert \mathbf{e}_{m,s} \rVert^{2}=1$, the solutions are given by
\begin{align}\label{optimal_EM_f_synthesis}
[\hat{\mathbf{f}}_{s}]_m
=
\sqrt{\varrho_i}
\lVert\mathbf{g}_{m}^{(r)}(\boldsymbol{\theta}_i)\rVert
,\ 
\hat{\mathbf{e}}_{m,s}
=
\frac{\mathbf{g}_{m}^{(r)}(\boldsymbol{\theta}_i)^{*}}{\lVert \mathbf{g}_{m}^{(r)}(\boldsymbol{\theta}_i)\rVert}
.
\end{align}

\subsection{Robust Codebook Design}
In the previous subsection, a low-complexity joint digital and \ac{em} codebook was derived from the solution of \eqref{opt_prob_represent} by minimizing the average \ac{crb}, assuming perfect knowledge of the channel parameters in \eqref{def:eta}. In practice, these parameters are unknown and must be estimated, motivating the need for a robust codebook design. To this end, we sample $L$ 2D-\acp{aod} $\{\boldsymbol{\psi}_\ell\}_{\ell=1}^{L}$ that span the angular uncertainty region and construct a codebook of $N_t = 3L$ codewords $\widehat{\mathbf{w}}_r$ based on \eqref{eq:w_opt_synthesis}:
\begin{equation}\label{def:synthesis_codebook_w}
\boldsymbol{\mathcal{W}}
=
\bigg\{
\sqrt{\delta_{3\ell -2}}\,
\widehat{\mathbf{w}}_1(\boldsymbol{\psi}_\ell),
\sqrt{\delta_{3\ell -1}}\,
\widehat{\mathbf{w}}_2(\boldsymbol{\psi}_\ell),
\sqrt{\delta_{3\ell}}\,
\widehat{\mathbf{w}}_3(\boldsymbol{\psi}_\ell)
\bigg\}_{\ell=1}^{L}.
\end{equation}
The corresponding digital and \ac{em} precoders are recovered from \eqref{optimal_EM_f_synthesis}. In \eqref{def:synthesis_codebook_w}, the power allocation coefficients $\{\delta_s\}$ are optimized following an approach similar to \cite{Fadakar2026Hybrid}, with details omitted due to space constraints.

\subsection{Proposed Channel Estimation Approach}
In this subsection, we present the proposed method for estimating the \acp{ecsi} matrices. 
Conventional \ac{ls} and subspace-based approaches \cite{Ying2024Reconfigurable} become computationally expensive when the system dimensions $M$, $Q$, $N_s$, or $I$ are large, and subspace methods additionally require a large number of samples for accurate \ac{scm} estimation. 
Motivated by the observation that wireless channels are often sparse in the delay-angle domain, particularly for large arrays and/or wide bandwidths, we estimate the channel parameters $\{\tau_i, \boldsymbol{\theta}_i, \alpha_i\}_{i=0}^{I-1}$, from which the \ac{ecsi} matrices are reconstructed using \eqref{eq:y-def-simp}.

The proposed approach consists of two stages. In Stage~1, the \ac{omp} algorithm is employed to obtain coarse estimates of the channel parameter vector $\boldsymbol{\eta}$ (see \eqref{def:eta}). By iteratively resolving propagation paths, this stage yields initial estimates of the delays, 2D-\acp{aod}, and complex gains, denoted by $\{\hat{\tau}_i, \hat{\boldsymbol{\theta}}_i, \hat{\alpha}_i\}_{i=0}^{\widehat{I}-1}$. In Stage~2, these estimates are refined using the \ac{sage} algorithm \cite{Fessler1994SAGE,Fleury1999SAGE}, which sequentially updates the parameters until convergence.

\section{Simulations}
This section evaluates the proposed codebook design and channel estimation performance for \ac{er-fas}-assisted mmWave systems via numerical simulations. The default system parameters are summarized in Table~\ref{tab:sys-params}. We consider $I=4$ propagation paths, including one \ac{los} path and three single-scatterer \ac{nlos} paths, with corresponding delays and 2D-\acp{aod} listed in Table~\ref{tab:sys-params}. The \ac{er-fas} employs a \ac{upa} with $M_1=M_2=7$. Based on the specified uncertainty regions for the \ac{ue} and scatterers, the elevation and azimuth domains are uniformly sampled with step sizes $d_\theta=d_\phi=1.8/M_1$~rad, yielding $L=9$ 2D-\acp{aod} and $N_t=3L=27$ codewords. 
The transmit power $P$ is adjusted to achieve different \ac{snr} levels on the \ac{los} path, defined as $\mathrm{SNR}=P\rho_0^2/(N_0B)$ \cite{Fadakar2026Hybrid}, and \ac{lmr}\cite{fadakar2024multi} is set to $\mathrm{LMR}=0\,\mathrm{dB}$. 
Performance is assessed using \ac{rmse} and \ac{crb} metrics, and optimal power allocation is applied in all simulations.
Finally, the weights $\delta_u$ in \eqref{opt_prob_represent} are selected to normalize the disparate scalings of the channel parameters by expressing delays in nanoseconds and \acp{aod} in degrees. The weights associated with the channel gains are set to zero, as these gains are obtained in closed form via \ac{ls} given the remaining parameters and therefore do not contribute additional information \cite{fadakar2025mutual, Fadakar2026Hybrid}.

\begin{figure}[!t]
\centering
\resizebox{\columnwidth}{!}{
\begin{tikzpicture}

\definecolor{col1}{RGB}{0, 0, 255}    
\definecolor{col2}{RGB}{230, 159, 0}   
\definecolor{col3}{RGB}{86, 180, 233}  
\definecolor{col4}{RGB}{204, 0, 0}    
\definecolor{col5}{RGB}{240, 228, 66}  
\definecolor{col6}{RGB}{143, 0, 143}   
\definecolor{col7}{RGB}{0, 0, 0}      
\definecolor{col8}{RGB}{245, 128, 180} 
\definecolor{col9}{RGB}{102, 51, 0}    
\definecolor{col10}{RGB}{153, 153, 153} 

\definecolor{darkgray176}{RGB}{176,176,176}
\definecolor{darkorange25512714}{RGB}{255,127,14}
\definecolor{lightgray204}{RGB}{204,204,204}
\definecolor{steelblue31119180}{RGB}{31,119,180}

\begin{axis}[
height=7cm,
width=12cm,
legend cell align={left},
legend style={
font=\fontsize{8}{10}\selectfont,
fill opacity=0.8, 
draw opacity=1, 
text opacity=1, 
draw=lightgray204
},
log basis y={10},
tick align=outside,
tick pos=left,
title={},
x grid style={darkgray176},
xlabel={SNR [dB]},
xmajorgrids,
xmin=-16.5, xmax=16.5,
xminorgrids,
xtick style={color=black},
y grid style={darkgray176},
ylabel={Average Delay Error [ns]},
ymajorgrids,
ymin=0.013054050774881, ymax=26.9568284209465,
yminorgrids,
ymode=log,
ytick style={color=black}
]

\addplot [semithick, col1, mark=o, mark size=3, mark options={solid}]
table {%
-15 15.0086231948712
-10 10.18116300620513
-5 0.395256982235471
0 0.2136875985257183
5 0.1124785644863325
10 0.065214785654823
15 0.038522479335719
};
\addlegendentry{RMSE $\tau$ (Traditional)}
\addplot [semithick, col1, dashed, mark=x, mark size=3, mark options={solid}]
table {%
-15 1.14044419951724
-10 0.641318902462174
-5 0.3606401214802
0 0.202802843829984
5 0.114044419951724
10 0.0641318902462173
15 0.0360640121480199
};
\addlegendentry{$\sqrt{\mathrm{CRB}}$ $\tau$ (Traditional)}

\addplot [semithick, col4, mark=o, mark size=3, mark options={solid}]
table {%
-15 7.14421062988167
-10 5.65454278541235
-5 0.21345484752641
0 0.108156185525159
5 0.059615428462684
10 0.03329859846843
15 0.01985618976268
};
\addlegendentry{RMSE $\tau$ (ER-FAS: $Q=4$)}
\addplot [semithick, col4, dashed, mark=x, mark size=3, mark options={solid}]
table {%
-15 0.584014511266265
-10 0.328415494195865
-5 0.184681604219146
0 0.103854098054874
5 0.0584014511266264
10 0.0328415494195865
15 0.0184681604219146
};
\addlegendentry{$\sqrt{\mathrm{CRB}}$ $\tau$ (ER-FAS: $Q=4$)}
\end{axis}

\end{tikzpicture}
}
\caption{
Average delay \ac{rmse} and $\sqrt{\mathrm{CRB}}$ over all paths versus \ac{snr}.
}
\label{fig:Avg_Delay}
\end{figure}
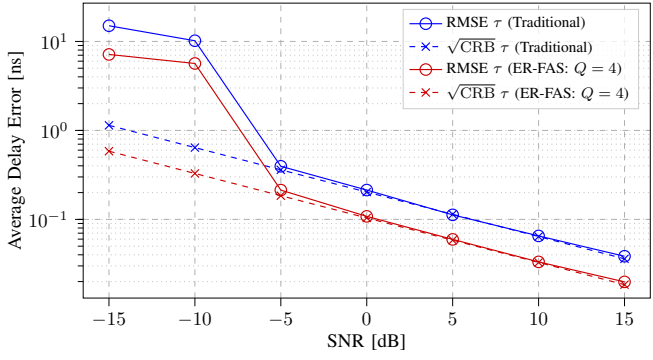

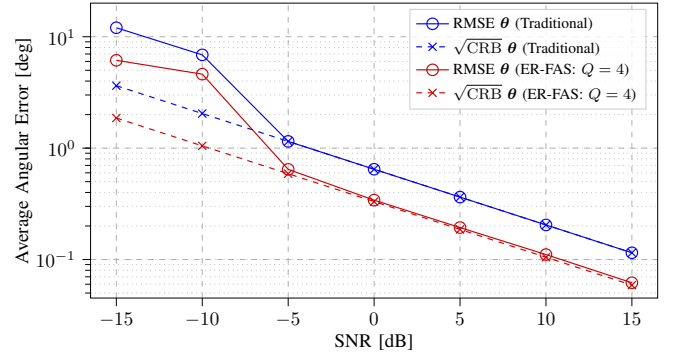
\begin{figure}[!t]
\centering
\resizebox{\columnwidth}{!}{
\begin{tikzpicture}

\definecolor{col1}{RGB}{0, 0, 255}    
\definecolor{col2}{RGB}{230, 159, 0}   
\definecolor{col3}{RGB}{86, 180, 233}  
\definecolor{col4}{RGB}{204, 0, 0}    
\definecolor{col5}{RGB}{240, 228, 66}  
\definecolor{col6}{RGB}{143, 0, 143}   
\definecolor{col7}{RGB}{0, 0, 0}      
\definecolor{col8}{RGB}{245, 128, 180} 
\definecolor{col9}{RGB}{102, 51, 0}    
\definecolor{col10}{RGB}{153, 153, 153} 

\definecolor{crimson2143940}{RGB}{214,39,40}
\definecolor{darkgray176}{RGB}{176,176,176}
\definecolor{darkorange25512714}{RGB}{255,127,14}
\definecolor{forestgreen4416044}{RGB}{44,160,44}
\definecolor{lightgray204}{RGB}{204,204,204}
\definecolor{steelblue31119180}{RGB}{31,119,180}

\begin{axis}[
height=7cm,
width=12cm,
legend cell align={left},
legend style={
font=\fontsize{8}{10}\selectfont,
fill opacity=0.8, 
draw opacity=1, 
text opacity=1, 
draw=lightgray204
},
log basis y={10},
tick align=outside,
tick pos=left,
title={},
x grid style={darkgray176},
xlabel={SNR [dB]},
xmajorgrids,
xmin=-16.5, xmax=16.5,
xminorgrids,
xtick style={color=black},
y grid style={darkgray176},
ylabel={Average Angular Error [deg]},
ymajorgrids,
ymin=0.0450702191005146, ymax=20.0355037329639,
yminorgrids,
ymode=log,
ytick style={color=black}
]

\addplot [semithick, col1, mark=o, mark size=3, mark options={solid}]
table {%
-15 12.04238099178738
-10 6.87625165152545
-5 1.14945133759073
0 0.646383988422588
5 0.363488428631381
10 0.204404564647928
15 0.114945133759072
};
\addlegendentry{RMSE $\boldsymbol{\theta}$ (Traditional)}
\addplot [semithick, col1, dashed, mark=x, mark size=3, mark options={solid}]
table {%
-15 3.6348842863138
-10 2.04404564647929
-5 1.14945133759073
0 0.646383988422588
5 0.363488428631381
10 0.204404564647928
15 0.114945133759072
};
\addlegendentry{$\sqrt{\mathrm{CRB}}$ $\boldsymbol{\theta}$ (Traditional)}

\addplot [semithick, col4, mark=o, mark size=3, mark options={solid}]
table {%
-15 6.1430259916538
-10 4.6151632685632
-5 0.64457792125458
0 0.341258785125641
5 0.193548781256784
10 0.11078545312587
15 0.06184865894152
};
\addlegendentry{RMSE $\boldsymbol{\theta}$ (ER-FAS: $Q=4$)}
\addplot [semithick, col4, dashed, mark=x, mark size=3, mark options={solid}]
table {%
-15 1.86307200159621
-10 1.04768237830265
-5 0.589155096993287
0 0.331306257985855
5 0.186307200159621
10 0.104768237830265
15 0.0589155096993289
};
\addlegendentry{$\sqrt{\mathrm{CRB}}$ $\boldsymbol{\theta}$ (ER-FAS: $Q=4$)}

\end{axis}

\end{tikzpicture}
}
\caption{
Average 2D-\ac{aod} \ac{rmse} and $\sqrt{\mathrm{CRB}}$ over all paths versus \ac{snr}.
}
\label{fig:Avg_AOD}
\end{figure}

Two array configurations are considered: a traditional non-reconfigurable array and the proposed \ac{er-fas} with $Q=4$ basis functions. 
Fig.~\ref{fig:Avg_Delay} and Fig.~\ref{fig:Avg_AOD} present the corresponding average \ac{rmse} and \ac{crb} over all propagation paths, obtained from $500$ Monte-Carlo realizations per \ac{snr}. The results demonstrate that the proposed \ac{er-fas} consistently outperforms the conventional array, highlighting its improved power efficiency.

\begin{table}[ht]
\caption{\label{tab:sys-params} System parameters}
\centering
\fontsize{12}{10}\selectfont 
\resizebox{\columnwidth}{!}{
\begin{tabular}{ |l|l|  }
\hline
Default System Parameters and
Symbol
&
\textbf{Value}
\\
\hline
Carrier frequency $f_c$ & $30\,\mathrm{GHz}$
\\
Noise PSD $N_0$ & $-173.855\,\mathrm{dBm}$
\\
Speed of light $c$ & $3\times 10^8\,\mathrm{m/s}$
\\
Subcarrier spacing $\Delta f$ & $200\,\mathrm{kHz}$
\\
Bandwidth $B$ & $200\,\mathrm{MHz}$
\\
\ac{ue} position $\mathbf{p}_u$ & $[45,5,1]^\mathsf{T}$
\\
Number of paths $I$ & $4$
\\
Scatterer positions $\{\mathbf{p}_i\}_{i=1}^{I}$
& 
$\mathbf{p}_1=[33,-7,1]^\mathsf{T}$, 
$\mathbf{p}_2=[42,2,8]^\mathsf{T}$, 
$\mathbf{p}_3=[35,5,5]^\mathsf{T}$
\\
Delays of paths $\{\tau_i\}_{i=0}^{3}$ (in ns) & 
$151.51,\ 169.80,\ 167.80,\ 153.75$
\\
Elevation \acp{aod} of paths $\{\theta^{\text{el}}_i\}_{i=0}^{3}$ & 
$95^\circ,\ 96.8^\circ,\ 85.9^\circ,\ 90^\circ$
\\
Azimuth \acp{aod} of paths $\{\theta^{\text{el}}_i\}_{i=0}^{3}$ & 
$6.3^\circ,\ -12.0^\circ,\ 2.7^\circ,\ 8.1^\circ$
\\
Number of \ac{shod} bases 
$Q$ &  $4$
\\
Uncertainty region & $30<x<50,-10<y<10$, $0<z<10$
\\
Elevation \ac{aod} bounds
$[\theta^{\text{el}}_\text{min}, \theta^{\text{el}}_\text{max}]$ 
& 
$[80.53^\circ, 99.46^\circ]$
\\
Azimuth a\c{aod} bounds 
$[\theta^{\text{az}}_\text{min}, \theta^{\text{az}}_\text{max}]$
&
$[-18.43^\circ, 18.43^\circ]$
\\
\ac{bs} position $\mathbf{p}_b$ & $[0,0,5]^\mathsf{T}\,\mathrm{[m]}$
\\
\ac{bs} \ac{upa} geometry $M_1$, $M_2$ & $7$, $7$
\\
Weighting coefficients $\sqrt{\delta_{(5i+1):(5i+5)}}$ & $180/\pi,180/\pi,10^9,0,0$
\\
\hline
\end{tabular}
}
\end{table}

\vspace{-0.3cm}
\section*{ACKNOWLEDGEMENT}
This work is supported 
by the National Science Foundation (Grants 2229535 and 2106602).

\bibliographystyle{IEEEtran}
\bibliography{Bib}

\end{document}